\documentclass[epj]{svjour}

\newcommand{\beq}{\begin{eqnarray}}
\newcommand{\eeq}{\end{eqnarray}}

\usepackage{graphicx}
\usepackage{fancyhdr}
\def\makeheadbox{{
 }}
\def\mytitle{My title} 
\def\myauthors{My name}  
\def\mytype{My type of session}
\def\mysession{My session}

\def\mytitle{MWT for Colliders} 
\def\myauthors{Mads Toudal Frandsen}    
\def\mytype{Contributed Talk}    
\def\mysession{Alternatives}

\begin{document}
\title{Minimal Walking Technicolor}
\subtitle{Setup for Collider Physics}
\author{Mads Toudal Frandsen
\thanks{\emph{Email:} toudal@nbi.dk}%
}                     
%
%
\institute{Niels Bohr Institute and University of Southern Denmark (SDU)}
%
\date{}
\abstract{
I report on our construction and analysis of the effective low energy Lagrangian for the Minimal Walking Technicolor (MWT) model \cite{Foadi:2007ue}. The parameters of the effective Lagrangian are constrained by imposing modified Weinberg sum rules and by imposing a value for the S parameter estimated from the underlying Technicolor theory. The constrained effective Lagrangian allows for an inverted vector vs. axial-vector mass spectrum in a large part of the parameter space. The effective Lagrangian presented here and more completely in \cite{Foadi:2007ue} is being implemented in Calchep and will be used to study the collider phenomenology of the MWT model in a later publication.
\PACS{
      {12.60.Nz}{Technicolor models}   \and
      {12.39.Fe}{Chiral Lagrangians}
     } 
} 
\maketitle
\section{Introduction}
\label{intro}
The Minimal Walking Technicolor (MWT) model was proposed by Sannino and Tuominen in \cite{Sannino:2004qp}. The model was extensively studied in  \cite{Dietrich:2005jn} and \cite{Dietrich:2005wk} in the light of electroweak precision measurements and  was shown to be a viable candidate for breaking the electroweak symmetry. In \cite{Foadi:2007ue} we have written down a comprehensive low energy effective theory for the MWT model as a first step in the study of the collider phenomenology, especially for LHC. For this purpose the model is currently being implemented in Calchep \cite{Belyaev}. 

\section{The MWT model}
\label{sec:the MWT model}
Here I briefly summarize the MWT model. For a more complete account of the model including Dark Matter candidates and gauge coupling unification see e.g.  \cite{Sannino:2004qp,Gudnason:2006yj,Gudnason:2006mk,Kainulainen:2006wq,Khlopov:2007ic}.
The model consists of an SU(2) technicolor gauge theory with two adjoint
Dirac technifermions.
The two adjoint fermions may be written as \beq Q^a=\left(\begin{array}{c}
U_L^a 
\\
D_L^a 
\\
-i\sigma^2 U_R^{*a} \\
-i\sigma^2 D_R^{*a}
\end{array}\right)
\ ,  \qquad a=1,2,3 \ ,\eeq with $a$ being the adjoint index of the technicolor SU(2). The left handed fields are arranged in three
doublets of the SU(2)$_L$ weak interactions in the standard fashion. 
The global symmetry is SU(4) and is assumed to break to the maximal diagonal subgroup $SO(4)$ via the condensate 
\beq \langle Q_i^\alpha Q_j^\beta
\epsilon_{\alpha \beta} E^{ij} \rangle =-2\langle \overline{U}_R U_L
+ \overline{D}_R D_L\rangle \ . \label{conde}
 \eeq
The electroweak gauge group is embedded into SU(4) such that this condensate  at the same time correctly breaks the electroweak symmetry \cite{Appelquist:1999dq}.
At this stage, however, the theory suffers from the Witten global anomaly, since it has an odd number of weak SU(2) doublets \cite{Witten:1982fp}. A simple way of avoiding the anomaly is to add a new weakly charged fermionic doublet which is a technicolor singlet. Schematically,
\beq L_L =
\left(
\begin{array}{c} N \\ E \end{array} \right)_L , \qquad N_R \ ,~E_R \
. \eeq Gauge anomalies cancel using the following generic hypercharge assignment:
\begin{eqnarray}
Y(Q_L)&=
\frac{y}{2} \ ,\qquad Y(U_R,D_R)=\left(\frac{y+1}{2},\frac{y-1}{2}\right) \ , \nonumber 
 \\
Y(L_L)&= -3\frac{y}{2} \ ,\qquad
Y(N_R,E_R)=\left(\frac{-3y+1}{2},\frac{-3y-1}{2}\right) \  .
\end{eqnarray}
$y$ can take any real value 
and one recovers the SM hypercharge
assignment for $y=1/3$.

\subsection{The Strong Sector of the MWT Model}
\label{sec:The Strong Sector of the MWT Model}
The strongly interacting sector of the MWT model is, in the ladder approximation, found to be close to the conformal window. More precisely, in this approximation an $SU(2)$ gauge theory of $N_F$ Dirac fermions in the adjoint representation of the gauge group enters the conformal window for $N_F\sim 2.1$ and looses asymptotic freedom for $N_f\sim 2.75$ \cite{Sannino:2004qp,Dietrich:2006cm}. The MWT model is therefore assumed to feature walking behavior (see \cite{Foadi:2007ue} for some of the original references on walking technicolor) of the gauge coupling   over a large range of energies. This results in a modified second Weinberg sum rule \cite{Appelquist:1998xf} and a reduced Peskin-Takeuchi S parameter \cite{Sundrum:1991rf,Appelquist:1998xf}. The walking dynamics further allows for the possibility of an inverted vector vs. axial-vector mass spectrum. The first study of these properties using lattice simulations were reported in 
\cite{Catterall:2007yx} and are currently under investigation
\cite{DelDebbio}. Being close to the the conformal phase transition might also render the Higgs mass of the MWT model light \cite{Dietrich:2005jn}.

\section{The Effective Theory}
\label{sec:The Effective Theory}
In \cite{Foadi:2007ue} we constructed an effective theory for the MWT model including composite scalars and vector bosons, their self interactions, and their interactions with the electroweak gauge fields and the standard model 
\newline
fermions. We start by describing the scalar sector but the focus here will be on the vector boson sector.

The scalar composites including the Higgs and the Goldstone bosons are assembled into a $4\times 4$ complex matrix $M$ with the quantum numbers of the techniquark bilinear, 
\begin{eqnarray}
M_{ij} \sim Q_i^\alpha Q_j^\beta \varepsilon_{\alpha\beta} \quad\quad\quad , {\rm with}\ i,j=1\dots 4. 
\end{eqnarray}
$M$ may be expanded using the set of broken generators $X$ in $SU(4)$ as 
\beq
M= \left[\frac{\sigma+i{\Theta}}{2} + \sqrt{2}(i\Pi^a+\widetilde{\Pi}^a)\,X^a\right]E \ .
\eeq
Under the SU(4) group $M$ transforms according to
\begin{eqnarray}
M\rightarrow uMu^T \ , \qquad {\rm with} \qquad u\in {\rm SU(4)} \ .
\end{eqnarray}
The electroweak gauge group $SU(2)_W\times U(1)_Y$ is embedded diagonally as a subgroup of a general $SU(4)$ gauge transformation matrix $G$ such that the electroweak covariant derivative for the $M$ matrix is
\begin{eqnarray}
D_{\mu}M =\partial_{\mu}M - i\,g \left[G_{\mu}M + MG_{\mu}^T\right]  \
, \label{covariantderivative}
\end{eqnarray}
where
\begin{eqnarray}
G_{\mu} & = &  W^a_\mu \ L^a + \frac{g^{\prime}}{g}\ B_\mu \ Y  \ , \ L^a=\left(\begin{array}{c}
 \frac{\tau^a}{2} \  0
\\
0 \quad  0
\end{array}\right) \ .
\label{gaugefields}
\end{eqnarray}
Here $g$ and $g'$ are the weak and hypercharge coupling respectively.
In terms of these fields the new Higgs Lagrangian is
\begin{eqnarray}
{\cal L}_{\rm Higgs} &=& \frac{1}{2}{\rm Tr}\left[D_{\mu}M D^{\mu}M^{\dagger}\right] - {\cal V}(M) + {\cal L}_{\rm ETC} \ ,
\end{eqnarray}
and  the potential reads
\begin{eqnarray}
{\cal V}(M) & = & - \frac{m^2}{2}{\rm Tr}[MM^{\dagger}] +\frac{\lambda}{4} {\rm Tr}\left[MM^{\dagger} \right]^2 \nonumber \\
& + &\lambda^\prime {\rm Tr}\left[M M^{\dagger} M M^{\dagger}\right] \nonumber \\
& - &  2\lambda^{\prime\prime} \left[{\rm Det}(M) + {\rm Det}(M^\dagger)\right] \ ,
\end{eqnarray}
${\cal L}_{\rm ETC}$ contains all terms which are generated solely by the ETC interaction. The renormalizable determinant terms explicitly break the U(1)$_{\rm A}$ symmetry, and give mass to the would-be Goldstone boson. While the potential has a (spontaneously broken) SU(4) global symmetry, the largest global symmetry of the kinetic term is \newline
SU(2)$_{\rm L}\times$U(1)$_{\rm R}\times$U(1)$_{\rm V}$.

\subsection{Vector Bosons}
\label{sec:Vector Bosons}
The composite vector bosons of the theory may be described by a $4\times 4$ hermitean traceless matrix $A^{\mu}$ with the quantum numbers of the vector bilinear  
\begin{equation}
A^{\mu,j}_{i}  \sim \ Q^{\alpha}_i  \sigma^{\mu}_{\alpha \dot{\beta}}  \bar{Q}^{\dot{\beta},j}
- \frac{1}{4} \delta_{i}^j Q^{\alpha}_k  \sigma^{\mu}_{\alpha \dot{\beta}} \bar{Q}^{\dot{\beta},k} \ .
\end{equation}
$A^{\mu}$ may be expanded using the complete set of generators  $T$ of SU(4), 
\begin{eqnarray}
A^\mu = A^{a\mu} \ T^a \ ,
\end{eqnarray}
and under the  SU(4) group $A^\mu$ transforms like
\begin{equation}
A^\mu \ \rightarrow \ u\ A^\mu \ u^\dagger \ ,\ \ \ {\rm with} \ u\in {\rm SU(4)} \ .
\label{vector-transform}
\end{equation}
To introduce the vector bosons in the effective Lagrangian framework one may formally gauge them and write a kinetic Lagrangian 
\begin{eqnarray}
{\cal L}_{\rm kinetic} = &-&\frac{1}{2}{\rm Tr}\Big[\widetilde{W}_{\mu\nu}\widetilde{W}^{\mu\nu}\Big] - \frac{1}{4}B_{\mu\nu}B^{\mu\nu} \nonumber
\\
&-&\frac{1}{2}{\rm Tr}\Big[F_{\mu\nu}F^{\mu\nu}\Big] 
+  m_A^2 \ {\rm Tr}\Big[C_\mu C^\mu\Big] \ .
\label{massterm}
\end{eqnarray}
$\widetilde{W}_{\mu\nu}$ and $B_{\mu\nu}$ are the field strength tensors for the electroweak gauge fields. The tilde on $W^a$ indicates  that the associated states are not yet the standard model weak triplets but mix with the composite vectors to form the ordinary $W$ and $Z$ bosons. $F_{\mu\nu}$ is the field strength tensor for the new SU(4) vector bosons,
\begin{eqnarray}
F_{\mu\nu} & = & \partial_\mu A_\nu - \partial_\nu A_\mu - i\tilde{g}\left[A_\mu,A_\nu\right]\ ,
\label{strength}
\end{eqnarray}
Up to dimension 4 the potential mixing $A^{\mu}$ and $M$ is 
\begin{eqnarray}
{\cal L}_{\rm M-C} & = & \tilde{g}^2\ r_1 \ {\rm Tr}\left[C_\mu C^\mu M M^\dagger\right]\nonumber \\
&+& \tilde{g}^2\ r_2 \ {\rm Tr}\left[C_\mu M {C^\mu}^T M^\dagger \right] \nonumber \\
& + & i \ \tilde{g}\ r_3 \ {\rm Tr}\left[C_\mu \left(M (D^\mu M)^\dagger - (D^\mu M) M^\dagger \right) \right] \nonumber \\
&+& \tilde{g}^2\ s \ {\rm Tr}\left[C_\mu C^\mu \right] {\rm Tr}\left[M M^\dagger \right] \ ,
\end{eqnarray}
and $\tilde{g}$ is the technicolor coupling. The vector field $C_\mu$ is defined by
\begin{eqnarray}
C_\mu \ \equiv \ A_\mu \ - \ \frac{g}{\tilde{g}}\ G_\mu (y) \ ,
\end{eqnarray}
This linear combination of the gauge fields transforms homogeneously under the electroweak symmetries.

\subsection{Constraining the Effective Theory}
\label{sec:Constraining the Effective Theory}
At this point the vector sector of the theory contains several new parameters. The technicolor coupling $\tilde{g}$, the common mass term $m_A$ and  the dimensionless parameters $r_1$, $r_2$, $r_3$, $s$. The latter  parameterize
the strength of the interactions between the composite scalars and
vectors in multiples of $\tilde{g}$. We can, however, constrain the effective theory in two ways. First, by imposing that the expression for the Peskin-Takeuchi $S$ parameter as computed in the effective theory be equal to the estimate for the $S$ parameter in the underlying theory. Secondly by imposing modified Weinberg sum rules to model the near conformal dynamics of the underlying theory at the effective Lagrangian level.

The first Weinberg sum rule is unaffected by the walking dynamics and implies the following relation in the effective theory:
\begin{equation}
F^2_V - F^2_A = F^2_{\pi}\ ,
\end{equation}
where $F^2_V$ and $F^2_A$ are the vector and axial vector decay
constants.
The second Weinberg sum rule is modified due to contributions from 
the near conformal region. It can be
expressed as \cite{Appelquist:1998xf}:
\begin{equation}
F^2_V M^2_V - F^2_A M^2_A = a\,\frac{8\pi^2}{d(R)}\,F_{\pi}^4 \ .
\label{2rule-2}
\end{equation}
$M_V$ ($M_A$) is the mass of the vector (axial vector) in the limit of no electroweak interactions and $a$ is a non-universal parameter expected to be positive and $O(1)$. $d(R)$ is the dimension of the representation of the underlying fermions. The unmodified second sum rule describing QCD-like dynamics with a running coupling constant has $a=0$.
In terms of the parameters of the effective theory we find:
\begin{eqnarray}
F^2_A &=& 2\frac{M^2_A}{\tilde{g}^2}(1 - \chi)^2 \ , \ F^2_{\pi} = v^2 ( 1 -\chi \, r_3 )\nonumber \\
F^2_V &=& 
\frac{2M^2_A}{\tilde{g}^2}=
\left(1 - \, \chi \frac{r_2}{r_3}\right)\, \frac{2M^2_A}{\tilde{g}^2} 
  \ ,\end{eqnarray}
where 
\begin{eqnarray}
\chi = \frac{v^2\tilde{g}^2}{2M^2_A} \, r_3 \ .
\label{S}
\end{eqnarray}
Hence the first WSR reads
\begin{eqnarray}
1+r_2 - 2 r_3 = 0 \ ,
\end{eqnarray}
while the second one reads
\begin{eqnarray}
 (r_2 - r_3) (v^2\tilde{g}^2 (r_2 + r_3) - 4 M^2_A) &-&\nonumber \\ 
a \frac{16\pi^2}{d(R)}  v^2\left( 1 - \chi \, r_3\right)^2 &=& 0\ .
\end{eqnarray}
The $S$ parameter expanded in $g/\tilde{g}$  is related to the parameter $r_3$ via:
\begin{eqnarray}
S= \frac{8\pi}{\tilde{g}^2}\, \chi \,(2- \chi)   \ .
\label{S}
\end{eqnarray}
It is instructive to consider the limit $\tilde{g} $ small while $g/\tilde{g}$ is still much smaller than one. To leading order in $\tilde{g}$ the second sum rule simplifies to:
\begin{eqnarray}
r_3 - r_2  = a \frac{4\pi^2}{d(R)}\frac{v^2}{M^2_A} \ ,
\end{eqnarray}
Together with the first sum rule we find:
\begin{eqnarray}
r_2 = 1 - 2 t \ , \qquad r_3 = 1 - t \ ,
\end{eqnarray}
with
\begin{eqnarray}
t= a \frac{4\pi^2}{d(R)}\frac{v^2}{M^2_A} \ .
\end{eqnarray}
The approximate $S$ parameter reads
\begin{eqnarray}
S= 8\pi\, \frac{v^2}{M^2_A} (1-t)   \ , 
\end{eqnarray}
so a small value of $a$ provides a large and positive $t$ rendering $S$ smaller than expected in a running theory. 

\section{Vector and Axial Vector Mass Spectrum}
\label{sec;Vector and Axial Vector Spectrum}
Here we show the values of $a$ and the vector and axial vector masses in the effective theory as function of $M_A$ and $\tilde{g}$ without making any approximations. We impose the first Weinberg sum rule and a conservative estimate of $S \sim 0.11$ given the dynamics of the underlying theory \cite{Appelquist:1998xf}. 
\begin{figure}[!htb]
\includegraphics[width=0.45\textwidth,height=0.30\textwidth,angle=0]{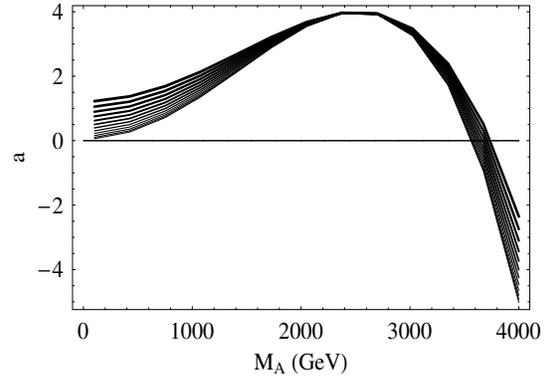}\caption{The parameter $a$ for $\tilde{g}$ varying from one (the thinnest curve)  to eight  (the thickest curve). Note that $a$ is expected to be positive or zero}
\label{fig:amamv}       
\end{figure}
\noindent
\noindent

In the limit of no electroweak interaction the masses of the vector and axial vector are given by the parameters $M_V$ and $M_A$.  
\begin{figure}[!htb]
\includegraphics[width=0.45\textwidth,height=0.25\textwidth,angle=0]{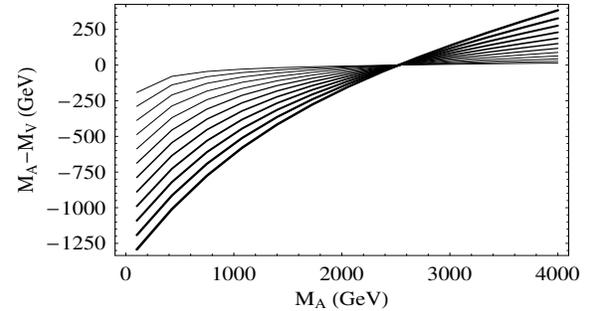}\caption{$M_A-M_V$ again as a function of $\tilde{g}$}
\label{fig:mamv}       
\end{figure}

\noindent
It is seen to be possible to have walking theories with light vector mesons and a small $S$ parameter given $a>0$ together with an axial meson lighter than its associated vector meson.  A degenerate spectrum allows for a small $S$ but with  relatively large values of $a$ and spin one masses around 2.5 TeV. We observe that $a$ becomes zero when the vector spectrum becomes sufficiently heavy, i.e.  we recover the running behavior for large masses of spin-one fields.

Finally we plot the physical masses of the vector ($\rho$) and the axial vector including electroweak corrections for a choice of $\tilde{g}=5$. As $\tilde{g}$ is lowered the vector moves towards the curve of the axial vector. Again the inverted mass spectrum is seen below $M_A\sim 2500$.
\begin{figure}[!h]
\includegraphics[width=0.45\textwidth,height=0.3\textwidth,angle=0]{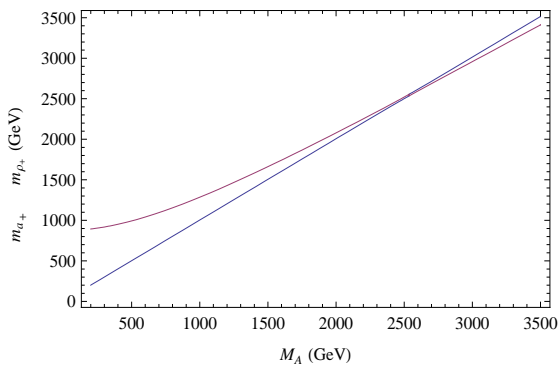}\caption{The vector $\rho$ (upper curve for small $M_A$) and axial vector masses for $\tilde{g}=5$ with electroweak interactions}
\label{fig:mrma}       
\end{figure}
\noindent

\section{Conclusions}
\label{sec:conclusions}
We have presented a comprehensive extension of the standard model at the effective Lagrangian level which embodies (minimal) walking technicolor theories and their interplay with the standard model particles. Our extension of the standard model features all of the relevant low energy effective degrees of freedom of the MWT model. These include scalars, pseudoscalars as well as spin one fields. Here we focused on the vector boson sector of the model.
The link with underlying strongly coupled gauge theories is achieved via the Weinberg sum rules taking into account the modification of the second sum rule due to walking. 
We have also analyzed the case in which the underlying theory behaves like QCD rather than being near an infrared fixed point. This has allowed us to gain insight into the spectrum of the spin one fields which is an issue of phenomenological interest. Finally we are implementing the MWT model in Calchep in order to efficiently study the phenomenology relevant for LHC. The outcome of this work will be presented in a later publication.

\section*{Acknowledgments}
\noindent
I would like to thank A. Belyaev, L. Del Debbio, D.D. Dietrich, R.Foadi, F. Sannino and T.Ryttov for collaboration and discussions on the material presented here.
The work of  M.T.F. is supported by the Marie Curie Excellence Grant under contract MEXT-CT-2004-013510.


\end{document}